\documentstyle[twocolumn,prl,aps,epsf]{revtex}
\begin{document}

\title{The interplay between $d$-wave superconductivity and
antiferromagnetic fluctuations: a quantum Monte Carlo study}

\author{F.F. Assaad}
\address{
   Institut f\"ur Theoretische Physik III, \\
   Universit\"at Stuttgart, Pfaffenwaldring 57, D-70550 Stuttgart, Germany. }

\maketitle

\begin{abstract}

We consider the  repulsive Hubbard model on a square lattice  with an 
additional term, $W$, which depends upon the square of a 
single-particle nearest-neighbor hopping. At half-band filling, 
constant $W$, we show that enhancing $U/t$ drives
the system from a $d$-wave superconductor to an antiferromagnetic Mott
insulator.  At zero temperature in the superconducting
phase,  spin-spin correlations  follow a powerlaw:
$ e^{-i \vec{r}  \cdot \vec{Q} } 
|\vec{r} | ^{-\alpha} $. Here  $\vec{Q} = (\pi, \pi)$ and $\alpha$ is in the
range $ 1  < \alpha < 2 $ and depends  upon
the coupling constants $W$ and $U$.  This results is reached on
the basis of  large scale quantum Monte-Carlo simulations on 
lattices up to 
$24 \times 24$, and is  shown  to be independent on the choice of the 
boundary conditions. 
We define a pairing (magnetic) scale by the temperature below which
{\it short range} $d$-wave pairing correlations  
(antiferromagnetic fluctuations) start growing. 
With  finite temperature quantum Monte Carlo 
simulations,  we demonstrate that both scales are identical over a large
energy range. Those results show the extreme 
compatibility  and close interplay  of antiferromagnetic fluctuations 
and $d$-wave superconductivity.

\end{abstract}


The understanding of the interplay between  $d$-wave superconductivity
and antiferromagnetism is a central issue for the  understanding  of the 
phase diagram of High-$T_c$ superconductors \cite{Mueller}. 
The aim of this work is to further study a model which shows a quantum
transition between a $d$-wave superconductor and an antiferromagnetic Mott 
insulator. It thus enables us to address the above question. 
The model we consider, has been introduced in Ref. \cite{Assaad_tUW_1}.
It is defined by: 
\begin{equation}
\label{tUW}
      H =  -\frac{t}{2} \sum_{\vec{i}} K_{\vec{i}}
          - W \sum_{\vec{i}} K_{\vec{i}}^{2} +
          U \sum_{\vec{i}}
         (n_{\vec{i},\uparrow}-\frac{1}{2})
         (n_{\vec{i},\downarrow} -\frac{1}{2})
\end{equation}
with the hopping kinetic energy
\begin{equation}
        K_{\vec{i}} = \sum_{\sigma, \vec{\delta}}
   \left(c_{\vec{i},\sigma}^{\dagger} c_{\vec{i} + \vec{\delta},\sigma} +
        c_{\vec{i} + \vec{\delta},\sigma}^{\dagger} c_{\vec{i},\sigma} \right).
\end{equation}
Here, $W \geq 0$,  $\vec{\delta} = \pm \vec{a}_x, \pm \vec{a}_y $, and 
$ n_{\vec{i},\sigma} = c^{\dagger}_{\vec{i},\sigma}  c_{\vec{i},\sigma}$
where $ c^{\dagger}_{\vec{i},\sigma}$  ($c_{\vec{i},\sigma}$) creates
(annihilates) an electron on site $ \vec{i} $ with z-component of spin
$\sigma$.
We impose twisted boundary conditions:
\begin{equation}
\label{Bound}
 c_{\vec{i} + L \vec{a}_x, \sigma } = \exp \left(2 \pi i \Phi/\Phi_0
\right) c_{\vec{i}, \sigma}, \; \;  c_{\vec{i} + L \vec{a}_y, \sigma }
= c_{\vec{i}, \sigma},
\end{equation}
with $\Phi_0 = h c / e$ the flux quanta and  $L$ the  linear length
of the square lattice.
The boundary conditions given by Eq. (\ref{Bound})
account for a magnetic flux threading a torus on which  the lattice is
wrapped.  One major advantage of the above model consists in the fact that at
half-band filling, the sign problem in the quantum Monte-Carlo  (QMC) method
may be  avoided. This statement is 
valid even for the above boundary conditions.  Hereafter, we will only consider
the half-filled case.

Previously we have considered the model  of Eq. (\ref{tUW})  at fixed values
of
$U$ and as a function of $W$. At $U/t = 4$ we have shown that the  quantum 
transition between the Mott insulator and $d$-wave superconductor 
occurs at $W_c/t \sim 0.3$ \cite{Assaad_tUW_1}.  
We have equally considered the model at finite
doping and given numerical evidence of the occurrence of a doping induced 
quantum transition between the Mott insulator and  $d$-wave  superconductor
\cite{Assaad_tUW_2}.
Here, we fix $W/t = 0.35$ and vary $U/t$ for the half-filled case.
The advantage of this choice of parameters is  that it 
provides us with  a 
{\it large} parameter range in the superconducting phase
where QMC simulations are extremely precise, thus allowing us to reach 
large lattices.  This allows us to reliably study the nature of the 
spin degrees of freedom in the superconducting state. 
We have used two QMC algorithms. i) The Projector
QMC algorithm which produces zero temperature results in a canonical
ensemble \cite{Koonin,Sandro}  and ii) the finite temperature grand canonical
QMC method  \cite{Hirsch85,White}. The application of those algorithms 
for the above model has been
discussed in reference \cite{Assaad_tUW_1}. Both algorithms, generate a 
systematic error proportional to $ (\Delta \tau)^2 $ where $\Delta \tau $
denotes the imaginary time step. To determine the exponent of the spin-spin
correlations we have  extrapolated  $\Delta \tau $ to zero.

\begin{figure}[ht]
\epsfxsize=10cm
\epsfxsize=7cm
\hfil\epsfbox{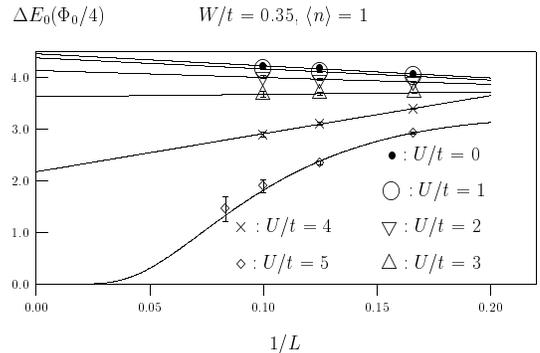}\hfil
\caption[]
{\noindent  $\Delta E_0 (\Phi_0/4) \equiv  E_0 ( \Phi_0/4 ) -
E_0 ( \Phi_0/2 ) $ as a function of  inverse linear size $L$.
For $U/t \leq 4$ the data is a consistent with
a $1/L$ form  which extrapolates
to a non-vanishing value in the thermodynamic limit. At $U/t = 5$ the data
may be fitted the SDW from $  L \exp ( -L/\xi ) $ thus signaling an
insulating state.
\label{suT0.fig} }
\end{figure}

We first concentrate of the charge degrees of freedom at zero temperature. 
To distinguish between a superconductor and insulator, we compute
the ground state energy as a function of the twist in the boundary
condition: $E_0(\Phi)$. 
For an insulator, the wave function is localized
and hence, an exponential decay of $ \Delta E_0(\Phi) \equiv E_0(\Phi) -
E_0(\Phi_0/2)$ as a function of lattice size is expected \cite{Kohn}.
In the Hartree-Fock  spin density wave (SDW)
approximation for the half-filled Hubbard model, one obtains
$\Delta E_0(\Phi)  = \alpha(\Phi) L \exp \left( -L/\xi \right)$ where
$\xi$  is the localization length of the wave function.
On the other hand, for a superconductor, $ \Delta E_0(\Phi) $
shows anomalous flux quantization: $ \Delta E_0(\Phi)$ is a
periodic function of $\Phi$ with period $ \Phi_{0}/2 $ and a
non vanishing energy barrier is to be found between the flux minima
\cite{Byers,Yang,Assaad1} so that $\Delta E_0(\Phi_0/4)$ remains finite as
$L \rightarrow \infty $.
Fig. (\ref{suT0.fig}) plots $ \Delta E_0(\Phi/4) $  versus $1/L$ 
at $W/t = 0.35$ and for various values of $U/t$. At values of 
$U/t \leq 4$, the data is consistent with  a $1/L$ form and scales to
a finite value. At $U/t = 5$, the data may be fitted to the above SDW
form. Thus, for values of $W/t = 0.35$, 
the data is consistent with the occurrence of a 
quantum transition between a superconductor  and a Mott insulator  at 
$4 < U_c/t < 5 $.

\begin{figure}[ht]
\epsfxsize=10cm
\epsfxsize=7cm
\hfil\epsfbox{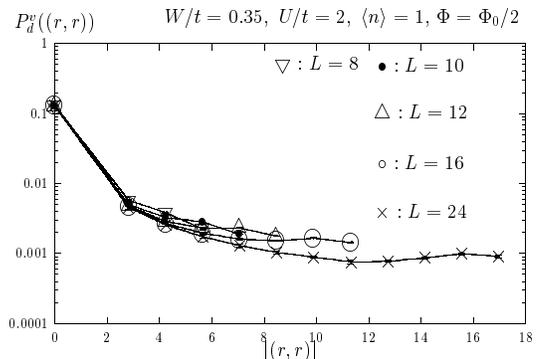}\hfil
\caption[]
{\noindent  Vertex contribution  to the $d$-wave pair field correlations.
Here, the temperature is set to $T=0$. \label{pair.fig} }
\end{figure}

In order to  determine the symmetry of the superconducting state,
we have computed equal time pairing correlations in the 
extended $s$-wave and
$d_{x^2 -  y^2}$  channels:
\begin{equation}
P_{d,s} (\vec{r}) = \langle \Delta_{d,s}^{\dagger}(\vec{r})
\Delta_{d,s}(\vec{0}) \rangle
\end{equation}
with
\begin{equation}
\Delta_{d,s}^{\dagger}(\vec{r})  =
\sum_{\sigma,\vec{\delta}}  f_{d,s}(\vec{\delta})
\sigma c^{\dagger}_{\vec{r},\sigma}
c^{\dagger}_{\vec{r} + \vec{\delta},-\sigma}.
\end{equation}
Here, $f_{s}(\vec{\delta}) = 1$ and $f_{d}(\vec{\delta}) = 1 (-1)$
for $\vec{\delta} = \pm \vec{a_x}$ ($\pm \vec{a_y})$.
The vertex contribution  to the above quantity is given by:
\begin{eqnarray}
\label{Pair_vertex}
& & P_{d,s}^{v} (\vec{r})  =  P_{d,s} (\vec{r}) 
-  \sum_{\sigma,\vec{\delta}, \vec{\delta}' }  f_{d,s}(\vec{\delta})
        f_{d,s}(\vec{\delta}') \\
 & & \left( \langle c^{\dagger}_{\vec{r},\sigma} c_{\vec{\delta}',\sigma}   \rangle
        \langle c^{\dagger}_{\vec{r}+\vec{\delta},-\sigma} c_{\vec{0},-\sigma}
        \rangle
 + \langle c^{\dagger}_{\vec{r},\sigma} c_{\vec{0},\sigma} \rangle
   \langle c^{\dagger}_{\vec{r}+\vec{\delta},-\sigma}
           c_{\vec{\delta}',-\sigma} \rangle \right). \nonumber
\end{eqnarray}
Per definition, $ P_{d,s}^{v}  (\vec{r}) \equiv 0 $ in the absence of
interactions.
Fig. \ref{pair.fig} plots $P_{d}^{v} (\vec{r}) $ along the diagonal of the 
lattice. We consider lattices ranging up to $L = 24$. 
For a fixed lattice size, one notices a plateau structure as   a function
of distance.  The extrapolation of this plateau value to the thermodynamic
limit is hard and the above criterion of flux quantization, proves to be 
a more efficient  method to conclude superconductivity
\cite{Note1}.
In comparison to the $d$-wave signal, the 
extended $s$-wave signal at {\it large } 
distances (data not shown) may not be distinguished 
from zero within our accuracy. 
The extended $s$-wave pair-field correlation dominate at short
distances. The data confirming this statement may be found in the table. 

The added $W$-term, contains no processes which explicitly 
favor $d$-wave superconductivity. On
the contrary, one would expect the $W$-term to favor extended 
$s$-wave symmetry, and this
shows up on short length scales. The fact that $d$-wave symmetry dominates
at long-range is a result of the underlying magnetic  structure.

\begin{figure}[ht]
\epsfxsize=10cm
\epsfxsize=7cm
\hfil\epsfbox{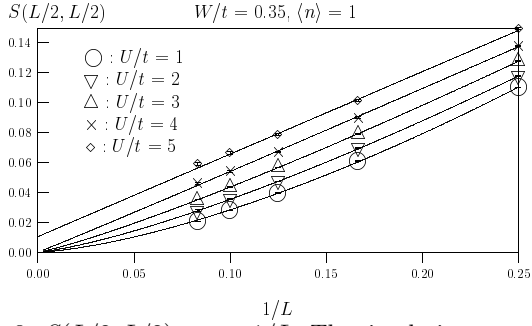}\hfil
\caption[]
{\noindent  $S(L/2,L/2)$ versus $1/L$. The simulations were carried out at
$\Phi/\Phi_0 =0$.   Here the  $\Delta  \tau$ systematic error is extrapolated
to zero and the temperature is set to $T=0$.
The solid lines are least square fits to the form: $L^{-\alpha}$.
The result of the fit is summarized in  Eq. (\ref{s_alpha}).
\label{spT0.fig} }
\end{figure}

\begin{figure}[ht]
\epsfxsize=10cm
\epsfxsize=7cm
\hfil\epsfbox{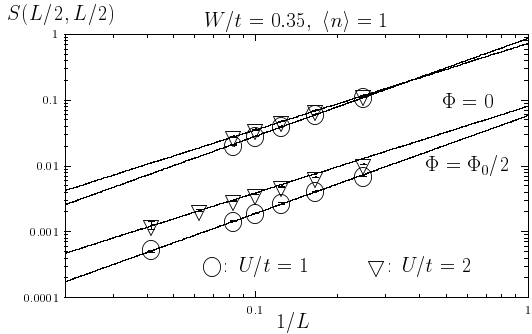}\hfil
\caption[]
{\noindent  $S(L/2,L/2)$ versus $1/L$ for different boundary conditions.
The solid line for the  $\Phi =0$ data is obtained from least square fit
to the form $L^{-\alpha}$. The same value of the exponent $\alpha$ was
used to fit the data at $\Phi = \Phi_0/2$.
Here the  $\Delta  \tau$ systematic error is extrapolated to zero, and
$T=0$.
\label{spin_ln.fig} }
\end{figure}

To study the spin degrees of freedom, we have computed equal time spin-spin
correlations 
\begin{equation}
  S(\vec{r}) = \frac{4}{3} \langle \vec{S} (\vec{r})   \vec{S} (\vec{0}) 
\rangle
\end{equation} 
where $ \vec{S} (\vec{r}) $ denotes the spin operator on site $\vec{r}$.
Fig.  \ref{spT0.fig}, plots  $S(L/2,L/2)$ versus $1/L$ where $L$ corresponds
to the linear size of the lattice. We consider periodic boundary conditions,
and various values of $U/t$. $W/t$ is constant and set to $W/t=0.35$. 
For values
of $U/t \leq 4 $ the data is consistent with a powerlaw decay:
\begin{eqnarray}
\label{s_alpha}
	S(L/2,L/2) \sim L^{-1.49} \; \; {\rm for } \; \; U/t=1  \nonumber \\
	S(L/2,L/2) \sim L^{-1.32} \; \; {\rm for } \; \; U/t=2 \nonumber \\
	S(L/2,L/2) \sim L^{-1.17} \; \; {\rm for } \; \; U/t=3  \nonumber \\
	S(L/2,L/2) \sim L^{-1.01} \; \; {\rm for } \; \; U/t=4.
\end{eqnarray}
At $U/t = 5$ extrapolation of the data leads to a finite staggered moment,
and thus the presence of antiferromagnetic long-range order.
The  statement in   Eq.  (\ref{s_alpha}) is surprising and deserves 
confirmation.  To cross-check the validity of the above equation, we 
demonstrate  numerically  that it is independent on the choice 
of the boundary condition. Fig.  \ref{spin_ln.fig} plots  $S(L/2,L/2)$ 
at $U/t  = 1 $ and $U/t = 2$  for $W/t = 0.35$.  At $\Phi =0 $, the 
solid lines corresponds
to least square fits to the form $ c L^{-\alpha} $.
We thus determine the exponent $\alpha$.   
The data at $\Phi = \Phi_0/2$ is consistent at large  distances
with the form $ \tilde{c}L^{-\alpha} $ thus showing that the exponent 
$\alpha $ is independent on the choice of the boundary. 
For the $\Phi = \Phi_0/2$
simulations we were able to reach  lattices up to $24 \times 24$.  
At $U/t = 2$ and for the boundary conditions set by $\Phi = \Phi_0/2$, 
$P_{d}^{v}(L/2,L/2)$  (see Fig. \ref{pair.fig})
decays slower than $S(L/2,L/2)$ (see Fig. \ref{spin_ln.fig}) thus 
confirming that the pairing correlations
are dominant \cite{Note1.5}.

\begin{figure}[ht]
\epsfxsize=10cm
\epsfxsize=7cm
\hfil\epsfbox{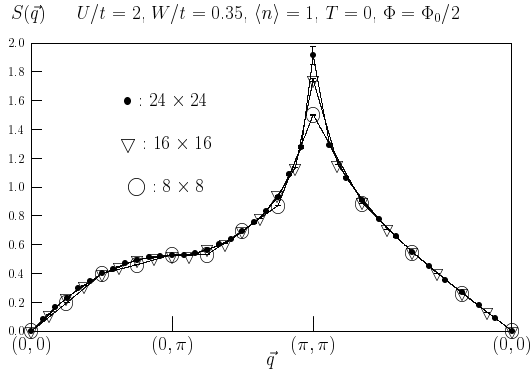}\hfil
\caption[]
{\noindent  Fourier transform of the equal time spin-spin correlations.
\label{sq.fig} }
\end{figure}

Having put Eq. (\ref{s_alpha}) on a numerically firm basis,
we now argue why it is
a surprising result. In two dimensions, correlation functions which decay
slower than $1/r^2$ lead to divergences in Fourier space. The
Fourier transform of the spin-spin correlations at $U/t = 2$ is plotted
in Fig. \ref{sq.fig}.  One sees a systematic increase of 
$S(\vec{Q} = (\pi,\pi) ) $ as a function of system size \cite{Note2}.
In contrast, a  mean field $d$-wave BCS calculation, 
yields spin-spin correlations which
decays as a powerlaw with $\alpha_{BCS} = 3.5$. This mean-field result
leads to a finite $S(\vec{Q}) $ in the thermodynamic limit. 
Another surprise comes from the dependence of the exponent $\alpha$ on 
the coupling constants, $U$ and $W$. This is a feature which occurs
in one dimensional quantum systems such as the $t-J$ or Hubbard models.

\begin{figure}[ht]
\epsfxsize=10cm
\epsfxsize=7cm
\hfil\epsfbox{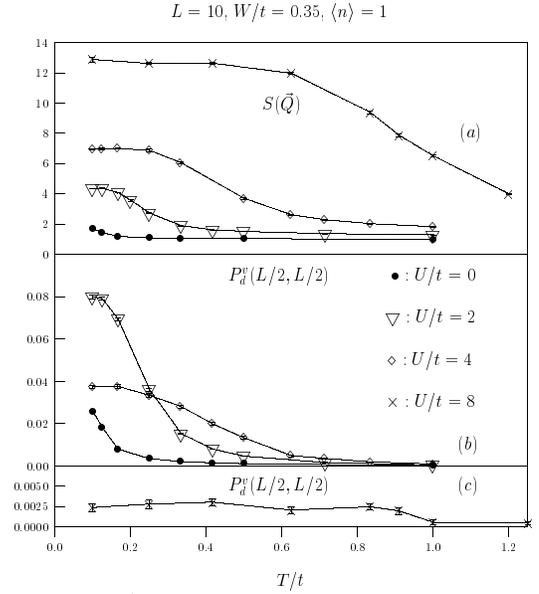}\hfil
\caption[]
{\noindent (a) $ S(\vec{Q} = (\pi,\pi) ) $ as a function of temperature, for
different values of $U/t$.
(b)-(c): $P_d^v (L/2,L/2) $ versus temperature.
The calculations presented in the figure were carried out at
$\Phi = 0$.
\label{TJTP.fig} }
\end{figure}

We now consider the model at finite temperatures with the use of the 
grand-canonical QMC algorithm. In Fig. \ref{TJTP.fig} (a) we plot S(Q)
for values of   $U/t$ ranging from  $U/t = 0$ to $U/t = 8$. We define
the magnetic scale, $T_J$, as the temperature scale where $S(Q)$ starts
growing. In Figs. \ref{TJTP.fig} (b) and (c) we consider the vertex
contribution to the $d$-wave paring correlations at  distance 
$\vec{R} = (L/2,L/2)$: $P^{v}_{d}(L/2,L/2)$. 
Here, we consider an $L=10$ lattice. At this distance the $s$-wave
pairing correlations are negligible. We  define the $d$-wave pairing scale, 
$T^{d}_p$, by the temperature scale at 
which $P^{v}_{d}(L/2,L/2)$ starts growing.
From the data, we conclude that  
\begin{equation}
	T^{d}_p \equiv  T_J
\end{equation}
That is,  antiferromagnetic fluctuations as well as $d$-wave pairing 
fluctuations occur hand in hand at the same energy scale.  
If  $ U/t > U_c/t $  ( $ U/t < U_c/t $ ) antiferromagnetic correlations 
($d$-wave pair-field correlations) will dominate at low temperatures. 
We note that in the large $U/t$ limit, $T_J$ should scale as
$t^2/U$.  That up to $U/t = 8$  we still do not see this
behavior, may be traced back  to the fact that the $W$-term enhances the 
band-width \cite{Assaad98}.
$T^{d}_p \equiv  T_J$ is 
a natural consequence of the  assumption of $SO(5)$ symmetry which unifies 
antiferromagnetism with $d$-wave superconductivity \cite{Zhang}.
The data, however, does not demonstrate the presence of this symmetry in
this model and further work is required.

In conclusion, we have considered aspects of the $t-U-W$ model defined in
Eq.(\ref{tUW}).  The model, at half-band filling, shows rich physics
and allows us to study the interplay between magnetism and $d$-wave 
superconductivity. We have pointed out the surprising nature of the spin-spin
correlations in the superconducting state: 
$ S(\vec{r}) \sim e^{-i \vec{r}  \cdot \vec{Q} }
|\vec{r} | ^{-\alpha} $, where  $\vec{Q} = (\pi, \pi)$,  $\alpha$ is in the
range $ 1 < \alpha < 2 $ and depends on
the coupling constants $W $ and $U$.
Those conclusions are based on large scale calculations  for system
sizes up to $24 \times 24 $.
We have equally shown that  the energy scales at which $d$-wave 
pairing  and antiferromagnetic
fluctuations occur are identical. The further understanding of spin 
and charge  dynamics,
as well as the doping of the model remains for further studies.

\section*{Acknowledgments}
M. Imada, D.J. Scalapino and M. Muramatsu are 
thanked for many instructive conversations.
The computations were carried out on the T3E of the HLRS, Stuttgart, as 
well as on the T90  and T3E of the HLRZ, J\"ulich.

\begin{table}[ht]
\begin{tabular}{|c|c|c|}
$ \vec{r} $ &  $P_{s}^{v} (\vec{r}) $ & $ P_{d}^{v} (\vec{r}) $ \\ \hline
$(0,0)$  &  $ 0.2950  \pm 0.0018 $  & $ 0.1304  \pm  0.0011 $  \\
$(0,1)$  &  $ 0.0932  \pm 0.0009 $  & $ 0.0238  \pm  0.0006 $  \\
$(0,2)$  &  $ 0.0076  \pm 0.0002 $  & $ 0.0252  \pm  0.0003  $  \\
\end{tabular}
\caption{ Short range  vertex contribution of pair-field correlations  in the
extended $s$- and $d$-wave channels. Here we consider an $L = 24 $ lattice
at $W/t = 0.35 $, $ U/t =2 $ and $\langle n \rangle = 1 $.  The
boundary conditions are set by $ \Phi = \Phi_0/2 $. The distance  
$ \vec{r} $ is in units of the lattice constant. }
\end{table}


\begin{thebibliography}{99}
\bibitem{Note1.5} At $U/t=2$ and $U/t = 1$, one may show numerically
that the vertex 
contribution to the $d$-wave pairing susceptibility grows quicker
than the vertex contribution to the staggered spin susceptibility
as a function of system size \cite{Assaad98}.
\end{thebibliography}
\end{document}